\documentclass[journal]{IEEEtran}
\ifCLASSOPTIONcaptionsoff
\newpage
\fi
\usepackage{graphicx}
\usepackage{amsmath}
\usepackage{epstopdf}
\usepackage{euscript}
\usepackage{algorithm,algorithmic}
\usepackage{algorithmic}
\usepackage{booktabs}
\usepackage{caption}
\usepackage{subfigure}
\usepackage{bm}
\usepackage{cite}
\usepackage{xcolor}
\usepackage{multirow}
\usepackage{verbatim}
\usepackage{amssymb}
\usepackage[numbers,sort&compress]{natbib}
\usepackage{multicol}
\usepackage{booktabs} 
\usepackage{setspace}
\begin{document}

\title{A CNN-LSTM Hybrid Framework for Wrist Kinematics Estimation Using Surface Electromyography}
\author{Tianzhe Bao,~\IEEEmembership{Student Member,~IEEE}, Syed Ali Raza Zaidi,~\IEEEmembership{Member,~IEEE}, Shengquan Xie,~\IEEEmembership{Senior Member,~IEEE}, Pengfei Yang,~\IEEEmembership{Member,~IEEE}, and Zhiqiang Zhang,~\IEEEmembership{Member,~IEEE}
	\thanks{\textit{(Corresponding author: Zhiqiang Zhang.)}}
	\thanks{Tianzhe Bao, Syed Ali Raza Zaidi, Shengquan Xie and Zhiqiang Zhang are with Institute of Robotics, Autonomous System and Sensing, School of Electrical and Electronic Engineering, University of Leeds, LS2 9JT, UK (e-mail: eltb@leeds.ac.uk; S.A.Zaidi@leeds.ac.uk; S.Q.Xie@leeds.ac.uk; Z.Zhang3@leeds.ac.uk).}
	\thanks{Pengfei Yang is with School of Computer Science and Technology, Xidian University, China (e-mail: pfyang@xidian.edu.cn).}}

\maketitle

\begin{abstract}
Convolutional neural network (CNN) has been widely exploited for simultaneous and proportional myoelectric control due to its capability of deriving informative, representative and transferable features from surface electromyography (sEMG). However, muscle contractions have strong temporal
dependencies but conventional CNN can only exploit spatial correlations. Considering that long short-term memory neural network (LSTM) is able to capture long-term and non-linear dynamics of time-series data, in this paper we propose a CNN-LSTM hybrid framework to fully explore the temporal-spatial information in sEMG. Firstly, CNN is utilized to extract deep features from sEMG spectrum, then these features are processed via LSTM-based sequence regression to estimate wrist kinematics. Six healthy participants are recruited for the participatory collection and motion analysis under various experimental setups. Estimation results in both intra-session and inter-session evaluations illustrate that CNN-LSTM significantly outperforms CNN and conventional machine learning approaches, particularly when complex wrist movements are activated.
\end{abstract}
\IEEEpeerreviewmaketitle\textbf{}

\section{Introduction}	
\IEEEPARstart{D}{uring} the past decades, there has been considerable attention given to surface electromyography (sEMG) in driving active prosthetic hands \cite{robertson2018effects}. To achieve intuitive myoelectric control, machine learning (ML) approaches, i.e. classifier-based pattern recognition (PR) and regression, have been extensively investigated in recent literature. Unlike PR-based methods which discriminate hand gestures in a discrete and sequential manner \cite{benatti2019online}, regression models focus on continuous wrist kinematics estimation \cite{shehata2018evaluating} and thus can promote simultaneous and proportional control in multiple degrees of freedoms (DoF). Several ML-based regression methods, including linear regression (LR), artificial neural network (ANN), kernel ridge regression, support vector regression (SVR) and random forest (RF), have been extensively exploited in both off-line simulations \cite{hahne2014linear,ameri2014support,ameri2014bagged,bakshi2018estimation,liu2019emg} and real-time prosthetic control \cite{hahne2018simultaneous}. However, ML techniques rely deeply on manual feature extraction \cite{khushaba2017framework}, i.e. feature engineering. Due to the stochastic nature of sEMG signals and serious crosstalk among muscles, useful information can be easily obscured in hand-crafted features. 

Deep learning (DL), particularly the convolutional neural network (CNN), is now providing a new perspective for feature learning/extraction via layer-by-layer processing \cite{zhou1702deep,zhao2019knowledge}. Promising results have been achieved in sEMG-based hand gesture recognition in the past few years. For instance, Park et al. presented a single stream CNN and evaluated the performance of DL learning via inter-subject estimations \cite{park2016movement}. Atzori et al. made a comprehensive comparison between CNN and several ML classifiers based on the NinaPro dataset \cite{atzori2016deep}. Du et al. presented an AdaBN-based deep domain adaptation scheme for inter-session recognition and conducted evaluations with two more public datasets (CSL-HDEMG and CapgMyo) \cite{du2017surface}. Wei et al. proposed a two-stage multi-stream CNN to learn the correlations between individual muscles \cite{wei2017multi}. Ding et al. proposed a parallel multiple-scale convolution architecture which exploited different size of kernel filters \cite{ding2018semg}. In addition, there are several pilot studies on regression-based wrist kinematics estimation. For instance, Ameri et al. investigated a CNN-based regression technique which outperformed a traditional SVR-based scheme in an online Fitts’ law test \cite{ameri2019regression}. Yang et al. presented several data-augmentation approaches for CNN in decoding 3-DoF wrist movements \cite{yang2018emg}. Although CNN is good at extracting spatial correlations of multi-channel sEMG signals, it inherently ignores the temporal information during continuous muscle contractions. 

Most recently, many researchers begin to implement the long short-term memory network (LSTM) for sEMG-based hand pose estimation. For example, Quivira et al. applied LSTM to build an accurate regression model for predicting hand joint kinematics from sEMG features \cite{quivira2018translating}. Teban et al. claimed that LSTM performed better than a non-recurrent ANN in replicating a non-linear mechanism of a real human hand \cite{teban2018recurrent}. He et al. combined LSTM with ANN to exploit both the dynamic and static information of sEMG \cite{he2018surface}. Ali et al. validated that a bidirectional LSTM with attention mechanism could outperform other tested recurrent neural networks (RNN) in sEMG-based hand gesture recognition \cite{samadani2018gated}. Despite that LSTM shows great effectiveness in capturing temporal dependencies based on learning contextual information from past inputs \cite{biswas2019cornet}, all those pilot studies have only applied conventional hand-crafted features rather than deep spatial features in their regression process. 

Inspired by advantages and limitations of CNN and LSTM, in this paper we propose a CNN-LSTM hybrid framework to combine deep feature extraction and sequence regression efficiently, so that the temporal-spatial correlations of sEMG can be fully exploited. With deep features extracted from CNN and then processed by LSTM, wrist kinematics in single/multiple DoFs can be reconstructed accurately. Compared with conventional CNN, CNN-LSTM is more robust to localized distortions along time. In this study, six healthy participants take part in experiments to perform a series of wrist movements. From experimental results it can be observed that CNN-LSTM outperforms CNN and conventional machine learning approaches significantly in intra-session/inter-session scenarios. The outperformance can be more evident when complex wrist movements are activated in multi-DoFs. 

The remainder of this paper is structured as follows. Section II describes the proposed hybrid framework, where the implementation of deep feature extraction and sequence regression are separately elaborated. Section III introduces experimental setups and Section IV presents estimation results in both intra-session and inter-session evaluations. In Section V the conclusion is drawn whilst the future work is presented.
  
\section{CNN-LSTM Hybrid Model} \label{CNN-LSTM}

As illustrated in Fig. 1, our CNN-LSTM model consists of two steps: the first step is to implement CNN for feature extraction and the second step is to construct LSTM for sequence regression. In the first step CNN is utilised to extract deep feature vector $\mathbf{f}$ from the sEMG matrix $\mathbf{X}$ which is constructed on a segment of multi-channel sEMG signals. In the second step, successive deep feature vectors are rearranged into a series of feature sequences, such as $\left[\mathbf{f}_{1}, \mathbf{f}_{2} \cdots \mathbf{f}_{k}\right]$, $\left[\mathbf{f}_{2}, \mathbf{f}_{3} \cdots \mathbf{f}_{k+1}\right]$, etc. The parameter $k$ is the number of feature vectors in a feature sequence, which denotes the time-steps in recurrent regression. A LSTM is built to convert $\left[\mathbf{f}_{1}, \mathbf{f}_{2} \cdots \mathbf{f}_{k}\right]$ into wrist angles $\left[y_{1}, y_{2}, \cdots y_{k}\right]$. In this study, we adopt the last output $y_{k}$ as the final observation of this sequence. In the following part we will elaborate the implementation of CNN and LSTM, together with the training process of each model.   

\subsection{CNN-based Deep Feature Extraction}
 
\subsubsection{Construction of sEMG Matrices}
Firstly, we use the sliding window method to split multi-channel sEMG into segments, and then signals in one segment are rearranged into a $1 \times L \times N$ matrix \cite{xia2018emg,hu2018novel}. Herein $L$ corresponds to the length of a sliding window and $N$ is in accordance with the number of sensor channels. By applying fast Fourier transform (FFT) on each channel, the spectrum-based sEMG matrix can be obtained as CNN inputs.

\subsubsection{CNN Architecture}
As illustrated in Fig. 2, the presented CNN consists of 4 convolutional blocks (Conv Block) and 2 fully connected blocks (FC Block). Each Conv Block has a convolutional layer, a batch normalization layer, a leaky ReLU layer, a max-pooling layer and a dropout layer. The convolution layer uses a kernel size of 3, a boundary padding of 1 and the stride of 1. There are 16 kernels in the 1$^{\text {st}}$ and 2$^{\text {nd}}$ Conv Block whilst 32 in the 3$^{\text {rd}}$ and 4$^{\text {th}}$ block. The batch normalization layer is attached to mitigate alternation made by convolutional layers \cite{santurkar2018does}. The leaky ReLU layer is used in case of the dying ReLU problem \cite{chitturi2018easily}. The max-pooling layer (a pool size of 3 and a stride of 1) is added for sub-sampling while a dropout layer is attached for regularization. In each FC Block, the batch normalization layer, leaky ReLU layer and dropout layer are added subsequently to the fully connected layer. There are 100 hidden units in the 1$^{\text {st}}$ FC Block and 20 in the 2$^{\text {nd}}$. Outputs of the 2$^{\text {nd}}$ FC Block will be utilized as the deep feature $\mathbf{f}$ for LSTM-based sequence regression.

\begin{figure}[t!] 
	\centering
	\setlength{\belowcaptionskip}{-0.1cm}   
	\includegraphics[width=3.5in]{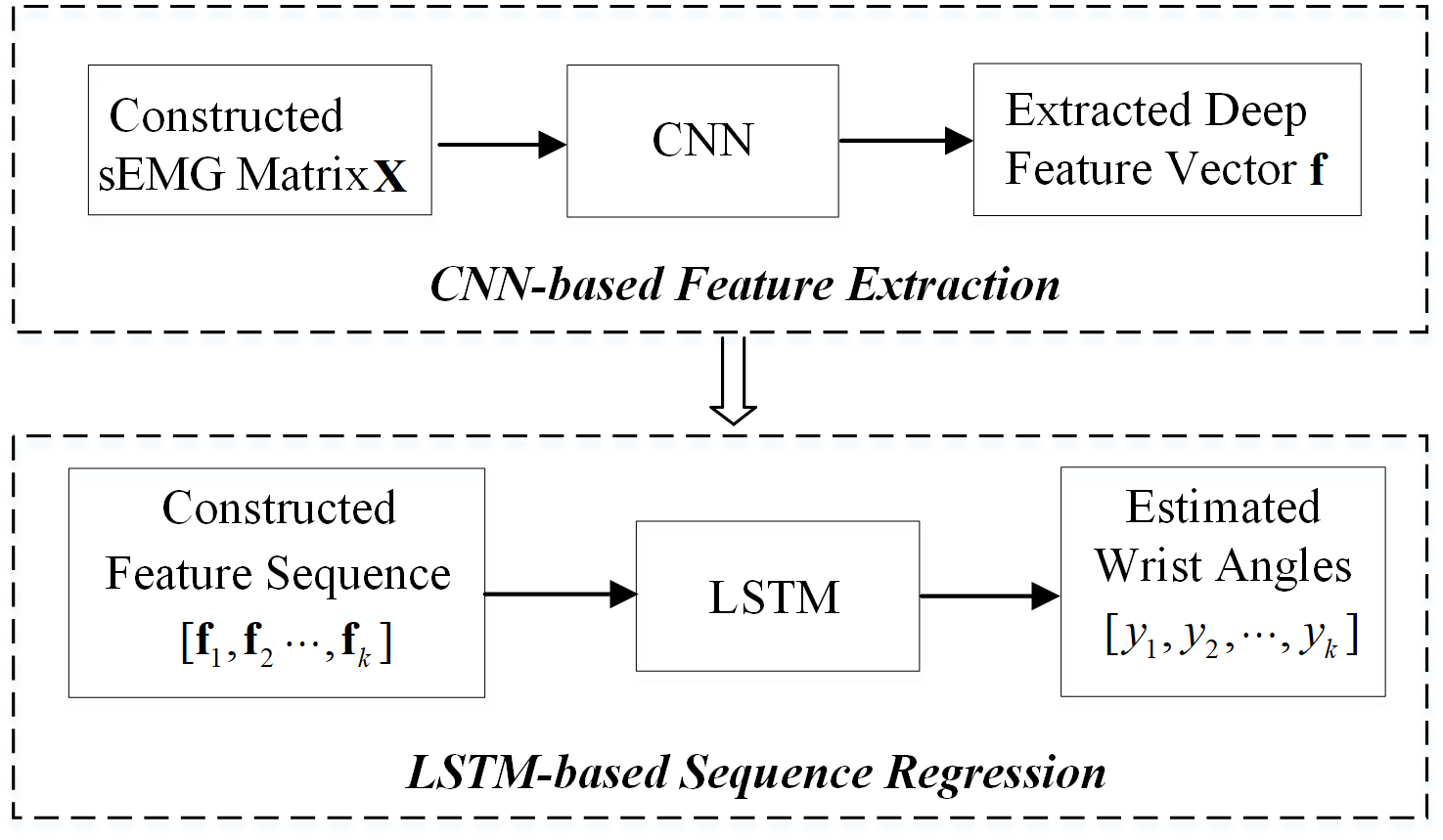}
	\caption{Block diagram of CNN-LSTM hybrid framework.}
	\label{CNN-LSTM}
\end{figure}  

\begin{figure}[b!] 
	\centering
	\includegraphics[width=3.5in]{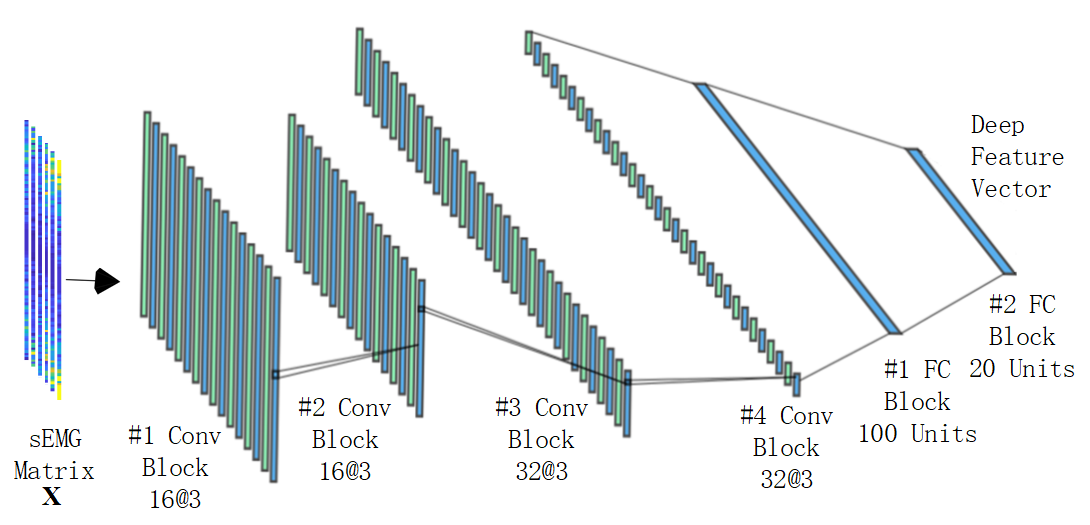}
	\caption{The single stream CNN architecture for deep feature extraction.}
	\label{CNN architectures}
\end{figure} 
   
\subsection{LSTM-based Sequence Regression}

\subsubsection{Topology of LSTM}
LSTM is a network designed to encode contextual information of a temporal sequence with feedback loops. It contains cycles that feed the network activations from a previous time-step to influence predictions at the current time-step \cite{sak2014long}. The unfolded chain structure of LSTM in an input sequence $\left[\mathbf{f}_{1}, \mathbf{f}_{2} \cdots \mathbf{f}_{k}\right]$ is illustrated in Fig. 3 \cite{bao2017deep}, where $\boldsymbol{h}_{j}$ ($j=1,2 \cdots k$) is the hidden state at time-step $j$ and $\boldsymbol{c}_{j}$ is the activation vector. In the recurrent regression, the LSTM unit uses previous state $\left(\boldsymbol{h}_{j-1}, \boldsymbol{c}_{j-1}\right)$ and current feature $\mathbf{f}_{j}$ to update current state $\left(\boldsymbol{h}_{j}, \boldsymbol{c}_{j}\right)$ and compute wrist angle $y_{j}$. In this way the historical information can be passed recursively in the whole loop of LSTM. 

\begin{figure}[t!] 
	\centering
	\includegraphics[width=3.5in]{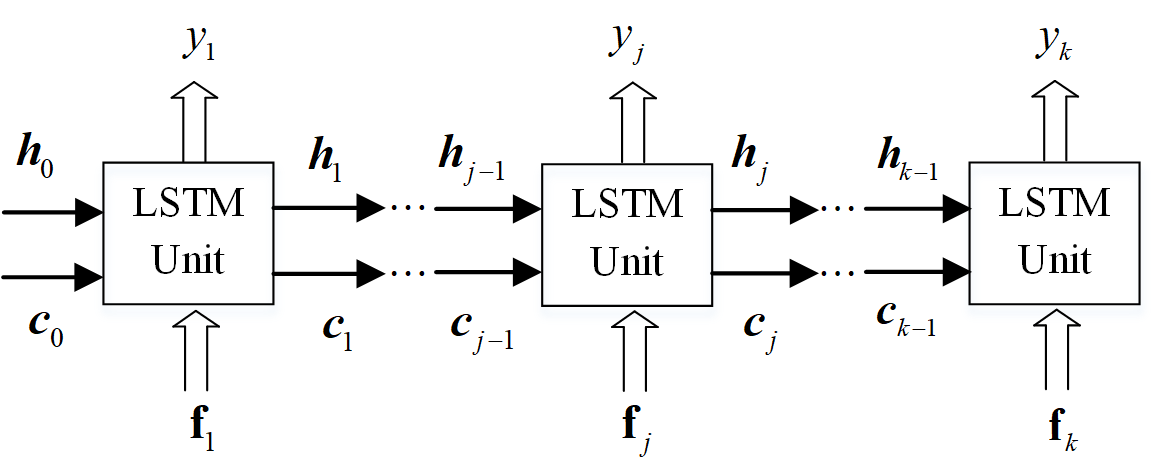}
	\caption{The unfolded chain structure of LSTM in time sequence with deep CNN features.}
	\label{LSTM-based sequence regression}
\end{figure} 

\subsubsection{Update of LSTM Units}
Basic elements of LSTM include an input gate to control activations for the memory cell, a forget gate to drop useless information of the past cell status, and an output gate to control the output activations for the ultimate state. The update of LSTM units at time-step ${j}$ can be described \cite{van2018automated}

\begin{equation}
\begin{split}
\label{Eq5}
&\boldsymbol{i}_{j}=\delta\left(\boldsymbol{W}_{i}\left[\boldsymbol{h}_{j-1,} \mathbf{f}_{j}\right]+\boldsymbol{b}_{i}\right)\\
&\boldsymbol{m}_{j}=\delta\left(\boldsymbol{W}_{m}\left[\boldsymbol{h}_{j-1,} \mathbf{f}_{j}\right]+\boldsymbol{b}_{m}\right)\\
&\boldsymbol{o}_{j}=\delta\left(\boldsymbol{W}_{o}\left[\boldsymbol{h}_{j-1,} \mathbf{f}_{j}\right]+\boldsymbol{b}_{o}\right)\\
&\boldsymbol{c_{j}}=\boldsymbol{i_{j}} \odot \tanh \left(\boldsymbol{W}_{c}\left[\boldsymbol{h}_{j-1,} \mathbf{f}_{j}\right]+\boldsymbol{b}_{c}\right)+\boldsymbol{m}_{j} \odot \boldsymbol{c}_{j-1}\\
&\boldsymbol{h}_{j}=\boldsymbol{o}_{j} \odot \tanh \left(\boldsymbol{c}_{j}\right)\\
&y_{j}=\boldsymbol{W}_{y} \boldsymbol{h}_{j}+\boldsymbol{b}_{y}
\end{split}
\end{equation}
where $\boldsymbol{i}_{j}$ is the input gate, $\boldsymbol{m}_{j}$ is the forget gate, $\boldsymbol{o}_{j}$ is the output gate, $\delta$ is the logistic sigmoid function, $\boldsymbol{W}$ is the weight matrix in each gate and layer, $\boldsymbol{b}$ is the corresponding bias vector and $\odot$ is the scalar product. The initial state $\left(\boldsymbol{h}_{0}, \boldsymbol{c}_{0}\right)$ will be settled after model training for subsequent predictions. 

\subsection{Training of CNN-LSTM} 
In this study we adopt the idea of separate training following the approach in reference \cite{wu2015modeling}. The main reason is that this strategy is much more computational efficient than the widely applied Long-term Recurrent Convolutional Networks (LRCNs) which trains CNN and LSTM jointly \cite{donahue2015long}. Besides, a component (CNN or LSTM) can be replaced without re-training the entire framework \cite{wu2015modeling}, which is more flexible in practical applications. Specifically, the tuning of CNN and LSTM is conducted in two subsequent steps. Firstly, a regression layer is attached to the presented CNN architecture to complete a supervised learning. In this step, the model inputs are sEMG matrices and observations are wrist angles. Secondly, deep feature vectors are extracted from the 2$^{\text {nd}}$ FC Block of CNN, based on which feature sequences are constructed to train LSTM for sequence regression.

\subsubsection{Training Setting of CNN}
Hyper-parameters of presented CNN are mainly identified referring to pilot studies in PR schemes \cite{atzori2016deep} and then determined via empirical manual tuning. As a general setting in this study, the network is trained in a 128-sized mini-batch as employed in  \cite{ameri2019regression} for 50 epochs by stochastic gradient descent with momentum (SDGM). The dynamic learning rate of CNN is 0.0001 in initialization and drops 90$\%$ after every 10 epochs. The slope scale is set as 0.1 in all leaky ReLU layers. The dropout rate in each dropout layer is 30$\%$. Other training strategies follow default settings in Matlab 2018b.

\subsubsection{Training Setting of LSTM}
In our study the time duration of a regression sequence is set to be 1 second. This achieves a trade-off between the information quantity of temporal dependencies and computational loads in practical implementation. LSTM is trained in a 64 sized mini-batch for 100 epochs via adaptive moment estimation (ADAM). The dynamic learning rate is initialised to be 0.001 and drops 90$\%$ after every 10 epochs. Since LSTM is prone to over-fitting more easily than conventional recurrent neural networks, herein only one LSTM layer with 50 hidden units is adopted. A dropout layer with 30$\%$ dropout rate is added for regularization. 

\section{Materials and Experimental Methods}
\subsection{Experiment Setup}
Approved by the Mathematics, Physical Science and Engineering joint Faculty Research Ethics Committee of University of Leeds, UK (reference MEEC 18-006), six healthy subjects (five males and one female, aged 24-30) took part in the experiment. The written informed consent was obtained from each subject before data collection. Following Fig. 4 (a), 12 bipolar electrodes were placed on the proximal portion of the forearm to collect sEMG signals in 6 channels. Reference electrodes were placed near the wrist. The inter-electrode distance in the proximal-distal direction was around 20 mm for reducing the crosstalk effect. 

As shown in Fig. 4 (b), in experiments participants were asked to perform four pre-defined wrist movement protocols. They were allowed to quit the experiments in case of any discomfort. The tested hand should be kept in a relaxing state to avoid muscle fatigue, with the upper limb supported vertically on the desk and the palm facing inside. All motions started from this rest position. Each protocol consisted of 3 sub-trials/sessions, and each session was composed of continuous wrist movements lasting around 3 minutes. A detailed description is reported in Table I. 

From Table I we can see that in P1-P3 only one DoF of the wrist motions was activated to complete single-DoF tasks. On the contrary, P4 aimed at multi-DoF tasks and all 3 DoFs were involved simultaneously. Obviously, P4 is naturally more complex and challenging compared with P1-P3 \cite{ameri2014real}, but it bears closer similarity with real-life movements \cite{bakshi2018estimation} and can speed up the training process. The frequency of sinusoidal contractions was around 0.1 Hz, meaning that a cycle of wrist rotation (such as rest-flexion-rest-extension-rest in P1) was about 10 seconds. 

In this study an attitude heading reference system (AHRS), composed of a tri-axial accelerometer, gyroscope and magnetometer, was utilized to obtain hand orientation \cite{madgwick2011estimation}. Wrist angles, which worked as the ground-truth in supervised learning, were calculated based on Euler angles from AHRS. Referring to Fig. 4 (b), both sEMG signals and wrist movements were recorded simultaneously with Shimmer wearable sensors \cite{burns2010shimmer} attached on the back of the testing hand. Sampling rates for accelerometer, gyroscope, magnetometer and sEMG were set as 100 Hz, 100 Hz and 75 Hz and 1024 Hz respectively. The online data streaming was implemented in a home-made software based on Shimmer Matlab Instrument Driver \cite{burns2010shimmer}. 
\begin{figure}[!t]
	\centering
	\includegraphics[width=3.5in]{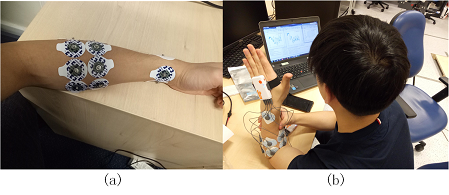}
	\caption{Experiment setup \cite{bao2019surface}. (a) Electrodes placement. (b) Data acquisition.}
	\label{CNN architectures}
\end{figure}  

\subsection{Data Pre-processing}
 In our experiments sEMG signals were processed using a 3$^{\text {rd}}$ order Butterworth high pass filter (20 Hz) to remove movement artifacts \cite{de2010filtering} and a low pass filter (450 Hz) to remove unusable high frequency noise \cite{micera2010control}. A notch filter with 50 Hz was also used to reduce power line noise. A Min-Max scaling was applied to normalize sEMG in each channel \cite{huang2018hybrid}. As for data segmentation, the analysis window was set to be 100ms with increment of 50 ms. Thus the size of sEMG matrix ($1 \times L \times N$) was $1 \times 101 \times 6$ in our experiments. Since the time duration of a feature sequence was set to be 1 second empirically, there were 18 time-steps in $\left[\mathbf{f}_{1}, \mathbf{f}_{2} \cdots \mathbf{f}_{k}\right]$, i.e. ${k=18}$.

\begin{table}[t]
	\centering 
	\caption {List of Performed Contractions} 
	\label{}
	\renewcommand{\arraystretch}{1.5}
	\footnotesize
	\begin{tabular}{p{1cm} p{3.2cm} p{3.3cm}}
		\hline
		Protocol & Description & Active DoF \\ \hline
		P1 & Sinusoidal contractions & Flexion-extension (F-E) \\ 
		P2 & Sinusoidal contractions & Pronation-supination (P-S) \\ 
		P3 & Sinusoidal contractions & Radial-ulnar deviation (R-U) \\ 
		P4 & Co-contractions of the wrist & F-E+P-S+R-U\\ \hline
	\end{tabular}
\end{table}

\subsection{Model Evaluation}
The analysis of sEMG-based wrist kinematics estimation was composed of intra-session and inter-session evaluations. To implement intra-session evaluations, the data in one session/trial of each protocol was split into four folds, where the first three were used for model training and the last for testing. To avoid data leakage, the splitting should be conducted before data pre-processing. In inter-session evaluations one whole session was used for model training and another session in the same protocol was used for testing. This method could better validate the model robustness against time-dependent changes of sEMG signals. 

The coefficient of determination ($R^{2}$) \cite{d2006control} was used as the metric to quantify the regression performance  

\begin{equation}
R^{2}=1-\frac{\operatorname{Var}\left(\boldsymbol{\alpha}^{d}-\mathbf{y}^{d}\right)}{\operatorname{Var}\left(\boldsymbol{\alpha}^{d}\right)}
\end{equation}
where $\boldsymbol{\alpha}^{d}$ are measured wrist angles by the sensor in $dth$ DoF and $\mathbf{y}^{d}$ are model estimations. According to Eq. (2), the numerator of $R^{2}$ is the mean squared error (RMSE) which is normalized by the variance of correct labels in the denominator \cite{hahne2014linear}. Compared with RMSE, $R^{2}$ is more robust to the numerical range of labels. $R^{2}$ at perfect estimation is equal to one, whilst a negative value means that estimation errors are larger than the variance of target values.
  
\section{Experimental Results} \label{EXPERIMENTAL RESULTS}
\subsection{Visual Exploration of sEMG Features}
Visual exploration is incredibly important in data-related problems since it allows intuitive analysis of the distributions or potential correlations between certain variables \cite{van2014accelerating}. In this section, t-Distributed Stochastic Neighbour Embedding (t-SNE) is utilized to project extracted CNN features (in testing sets) into two principal dimensions for visualization \cite{maaten2008visualizing}. For comparison, a widely applied temporal-spatial feature set \cite{phinyomark2013emg,muceli2011simultaneous} consisting of mean absolute value (MAV), root mean square (RMS), variance (VAR) and 4$^{\text {th}}$ order autoregressive coefficients (4$^{\text {th}}$ AR) are calculated. Scatter plots of projected sEMG features in P1 (F-E), P4 (F-E) and P4 (P-S) of intra-session evaluations are shown in Fig. 5, where the two axes represent two principal features, respectively. The angles of scatters (features) are reflected in parula colormap, with the pure yellow representing the positive maximal values in one DoF and pure purple for the negative maximum.

From Fig. 5 we can see that in each dataset the clustering of scatters projected from CNN features is significantly better than that of hand-crafted features. In the left part of each sub-figure, scatters with similar colour are gathering whilst those with different colours are highly distinguishable. On the contrary, scatters in the right one are overlapped heavily, even among the yellow ones and the blue ones. Compared with P1 (F-E), the clustering of scatters becomes worse for hand-crafted features in P4 (F-E). This deterioration becomes more evident in P4 (P-S), where distributions among scatters from CNN features become also ambiguous. A possible reason for the deterioration is that the crosstalk of sEMG can be quite serious in multi-DoFs tasks due to our forearm anatomy \cite{phinyomark2013emg}. Since muscle fibres of extensors and flexors are much thicker and also located in a more superficial layer of the forearm, information of other DoFs are easier to be buried in compounded sEMG.

\begin{figure}[!t]
	\centering
	\setlength{\belowcaptionskip}{-0cm}   
	\includegraphics[width=3.5in]{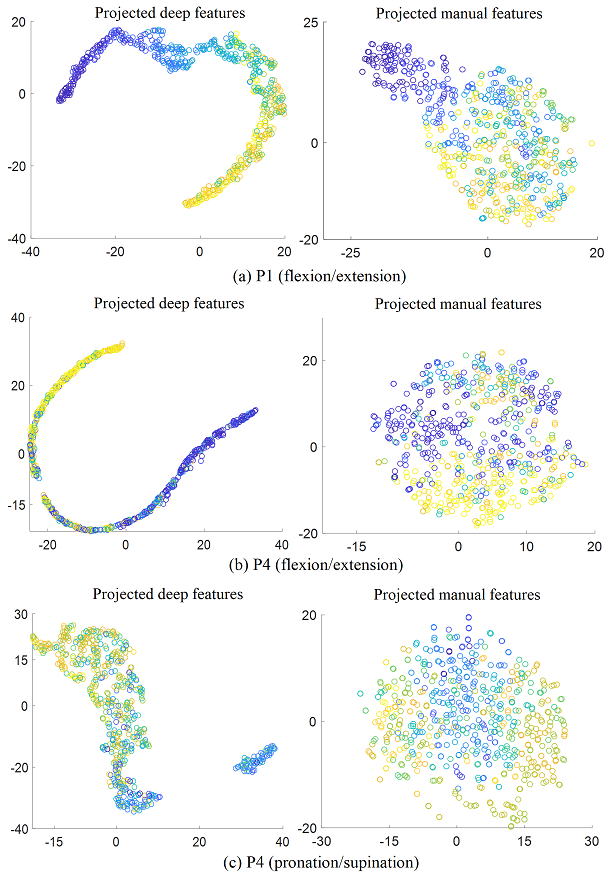}
	\caption{Distribution of CNN features and hand-crafted features in testing sets of Subject 5 after dimension reduction. Scatters in (a)-(c) correspond to features from P1 (F-E), P4 (F-E) and P4 (P-S), respectively.}
	\label{Visual Exploration} 
\end{figure}

\subsection{Intra-session Estimations in Single-DoF Tasks}
Fig. 6 shows wrist angles captured by AHRS system in P1-P3 of Subject 5 together with estimations of CNN and CNN-LSTM. As illustrated in the figure, trajectories reconstructed by CNN-LSTM (in red) are smoother and much closer to the ground-truth (in blue) than CNN trajectories (in yellow) in all tasks. In fact, during continuous muscle contractions there are supposed to be strong temporal-dependencies in the produced sEMG signals. Thus it is reasonable to consider sEMG as time-series data in regression tasks. Different from conventional CNN that only focuses on spatial correlations among channels/electrodes, the history information of successive deep feature vectors in a sequence $\left[\mathbf{f}_{1}, \mathbf{f}_{2} \cdots \mathbf{f}_{k}\right]$ is further exploited by CNN-LSTM, which improves estimation accuracies significantly. Another interesting result is that the estimated trajectories of both CNN and CNN-LSTM in P1 are better than their corresponding results in P2 and P3. This is consistent with the visual exploration, where feature scatters in flexion and extension are much more distinguishable than those in the other two DoFs. 
\begin{figure}[!t]
	\centering
	\setlength{\belowcaptionskip}{-0cm}   
	\includegraphics[width=3.5in]{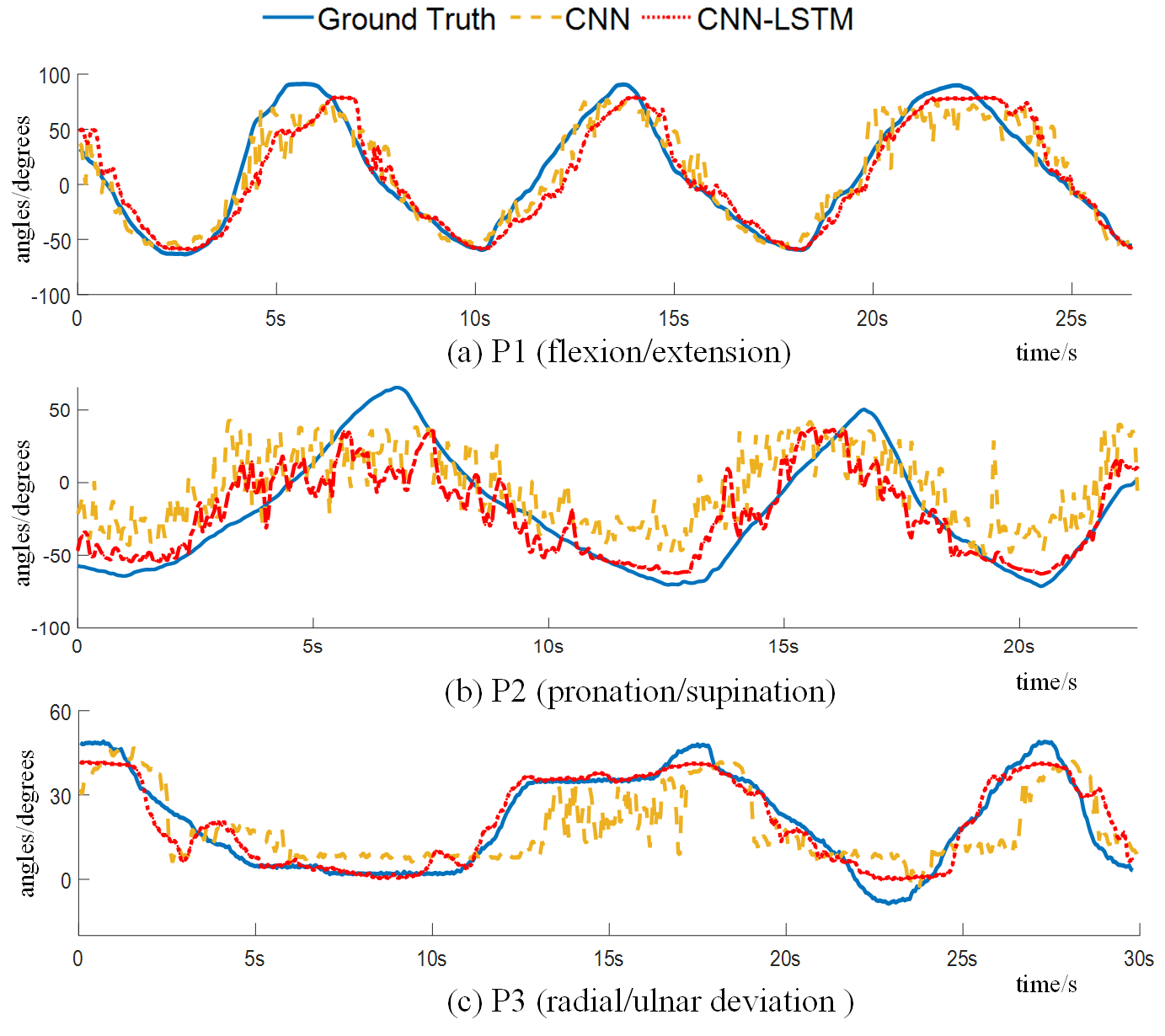}
	\caption{Wrist motions and intra-session estimations of CNN and CNN-LSTM for P1-P3 in Subject 5. (a) P1 (F-E). (b) P2 (P-S). (c) P3 (R-U).}
	\label{intra-session estimation}
\end{figure}

In this section two representative ML models, i.e. SVR and random forest (RF), are implemented to compare with DL techniques. SVR can project sEMG features into a higher dimensional space via kernel functions, whilst RF is currently the most popular ensemble learning technique. The outperformance of SVR and RF over other shallow models such as LR and ANN have been verified in pilot studies \cite{hahne2014linear,ameri2014bagged}. Consistent with visual exploration, MAV, RMS, VAR and 4$^{\text {th}}$ AR are working as hand-crafted features for ML models. To make a fair comparison between traditional ML models and the proposed CNN-LSTM framework, we have reduced the dimension of hand-crafted features to 20 using Principle Component Analysis (PCA). Following previous studies \cite{hahne2014linear}, a radial basis function (RBF) is adopted for SVR. Besides, the hyper-parameters of SVR and RF are optimized via the 5-fold inner cross-validation in each trial. 

\begin{table}[!t]
	\centering
	\caption {$R^{2}$ of SVR, RF, CNN and CNN-LSTM in Single-DoF Tasks (P1-P3) of Intra-session Evaluations} 
	\setlength{\tabcolsep}{3.0mm}
	\renewcommand\arraystretch{1.2}
	\footnotesize
	\begin{tabular}{c|l|lllc}
		\hline
		Subjects           & Protocols & SVR  & RF   & CNN  & CNN-LSTM               \\ \hline
		\multirow{3}{*}{1} & P1(F-E)  & 0.56 & 0.71 & 0.56 & \textit{\textbf{0.92}} \\ \cline{2-6} 
		& P2(P-S)  & 0.26 & 0.28 & 0.32 & \textit{\textbf{0.65}} \\ \cline{2-6} 
		& P3(R-U)  & 0.56 & 0.59 & 0.66 & \textit{\textbf{0.87}} \\ \hline
		\multirow{3}{*}{2} & P1(F-E)  & 0.60 & 0.69 & 0.66 & \textit{\textbf{0.85}} \\ \cline{2-6} 
		& P2(P-S)  & 0.37 & 0.48 & 0.45 & \textit{\textbf{0.56}} \\ \cline{2-6} 
		& P3(R-U)  & 0.22 & 0.25 & 0.31 & \textit{\textbf{0.64}} \\ \hline
		\multirow{3}{*}{3} & P1(F-E)  & 0.35 & 0.38 & 0.42 & \textit{\textbf{0.80}} \\ \cline{2-6} 
		& P2(P-S)  & 0.46 & 0.63 & 0.58 & \textit{\textbf{0.83}} \\ \cline{2-6} 
		& P3(R-U)  & 0.18 & 0.19 & 0.22 & \textit{\textbf{0.56}} \\ \hline
		\multirow{3}{*}{4} & P1(F-E)  & 0.35 & 0.40 & 0.41 & \textit{\textbf{0.75}} \\ \cline{2-6} 
		& P2(P-S)  & 0.17 & 0.16 & 0.21 & \textit{\textbf{0.46}} \\ \cline{2-6} 
		& P3(R-U)  & 0.40 & 0.48 & 0.43 & \textit{\textbf{0.88}} \\ \hline
		\multirow{3}{*}{5} & P1(F-E)  & 0.84 & 0.86 & 0.84 & \textit{\textbf{0.91}} \\ \cline{2-6} 
		& P2(P-S)  & 0.51 & 0.52 & 0.62 & \textit{\textbf{0.71}} \\ \cline{2-6} 
		& P3(R-U)  & 0.59 & 0.71 & 0.67 & \textit{\textbf{0.90}} \\ \hline
		\multirow{3}{*}{6} & P1(F-E)  & 0.71 & 0.76 & 0.74 & \textit{\textbf{0.91}} \\ \cline{2-6} 
		& P2(P-S)  & 0.21 & 0.30 & 0.36 & \textit{\textbf{0.64}} \\ \cline{2-6} 
		& P3(R-U)  & 0.36 & 0.32 & 0.40 & \textit{\textbf{0.69}} \\ \hline
	\end{tabular}
\end{table} 

Table II summarizes intra-session performances of SVR, RF, CNN and CNN-LSTM in P1-P3 of Subject 1-6. It can be inferred that the presented hybrid framework outperforms SVR, RF and CNN dramatically in all trials of all protocols. The outperformance of CNN-LSTM over other regression techniques can be extremely evident in some datasets, such as P2 and P3 in nearly all participants. Besides, the conventional single-stream CNN is in general comparable to SVR and RF in sEMG-based wrist kinematics estimation. This result is consistent with pilot studies in hand gesture recognition \cite{atzori2016deep}.
\subsection{Intra-session Estimations in Multi-DoF Tasks}
Different from single-DoF tasks (P1-P3), the multi-DoF task (P4) requires co-activations of 3 DoFs. Fig. 7 demonstrates the intra-session estimations of CNN and CNN-LSTM in P4 of Subject 5. In accordance with P1-P3, the reconstructed trajectories of CNN-LSTM are also closer to the ground-truth in all DoFs. As for $R^{2}$ values, CNN-LSTM reaches much higher scores than CNN, RF and SVR, indicating an evident improvement in model accuracy. $R^{2}$ values of each DoF in six subjects are listed in Table III. Same to P1-P3, performances of CNN, RF and SVR are in general close to each other. Consistent with results in visual exploration, deteriorations in estimation accuracies can be found in each DoF of P4 compared with those in P1-P3, indicating that the features of samples become harder to recognize.

\begin{figure}[!t]
	\centering
	\setlength{\belowcaptionskip}{-0cm}   
	\includegraphics[width=3.5in]{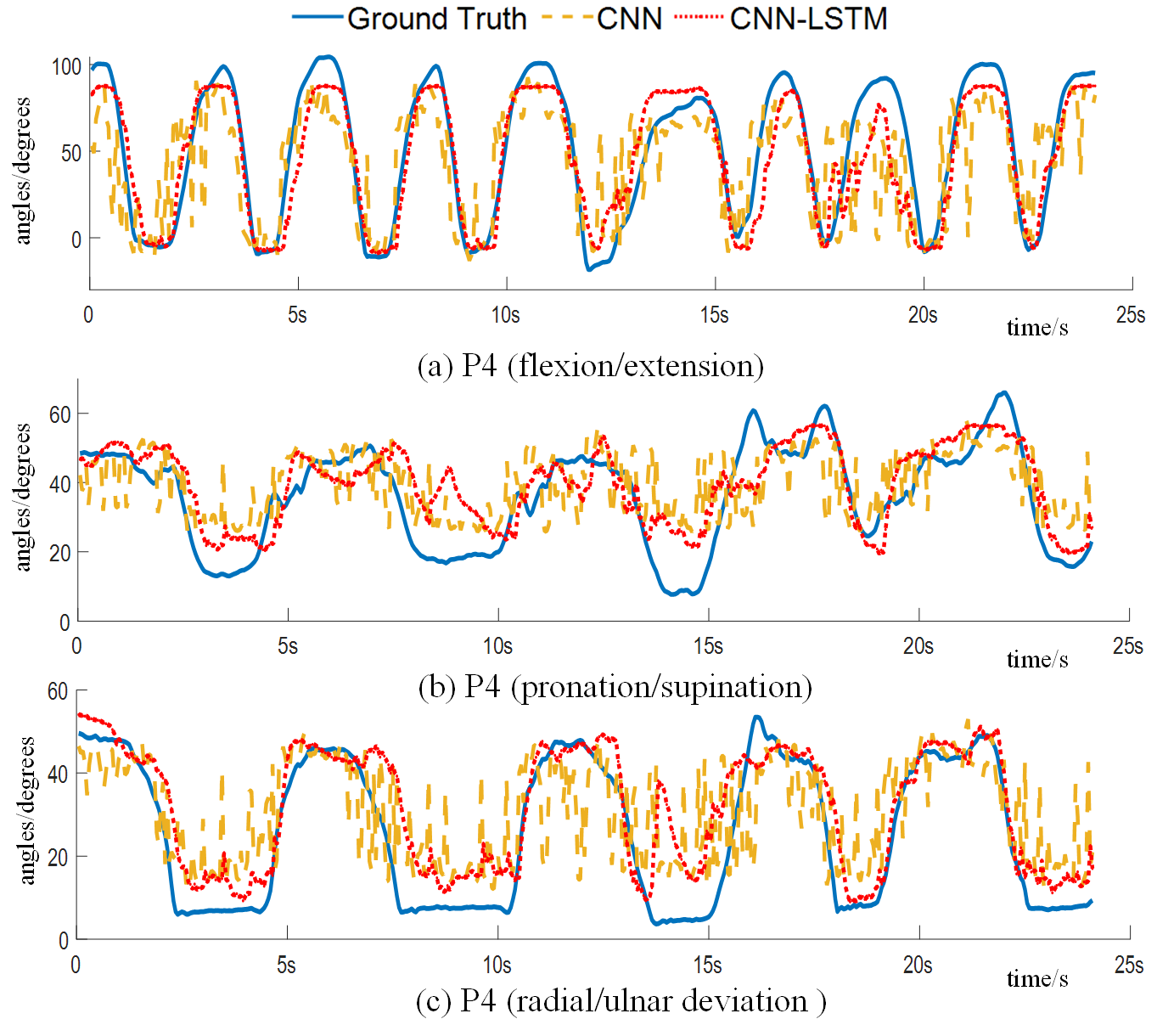}
	\caption{Wrist motions and intra-session estimations of CNN and CNN-LSTM in P4 of Subject 5. (a) P4 (F-E). (b) P4 (P-S). (c) P4 (R-U).}
	\label{intra-session estimations in P4}
\end{figure}

\begin{table}[!t]
	\centering
	\caption {$R^{2}$ of SVR, RF, CNN and CNN-LSTM in Multi-DoF Tasks (P4) of Intra-session Evaluations}
	\setlength{\tabcolsep}{3.0mm}
	\renewcommand\arraystretch{1.2}
	\footnotesize
	\begin{tabular}{c|l|lllc}
		\hline
		Subjects           & DoF & SVR  & RF   & CNN  & CNN-LSTM               \\ \hline
		\multirow{3}{*}{1} & F-E & 0.44 & 0.52 & 0.55 & \textit{\textbf{0.87}} \\ \cline{2-6} 
		& P-S & 0.30 & 0.31 & 0.30 & \textit{\textbf{0.58}} \\ \cline{2-6} 
		& R-U  & 0.40 & 0.38 & 0.39 & \textit{\textbf{0.69}} \\ \hline
		\multirow{3}{*}{2} & F-E  & 0.63 & 0.63 & 0.62 & \textit{\textbf{0.82}} \\ \cline{2-6} 
		& P-S  & 0.19 & 0.27 & 0.28 & \textit{\textbf{0.47}} \\ \cline{2-6} 
		& R-U  & 0.37 & 0.36 & 0.40 & \textit{\textbf{0.61}} \\ \hline
		\multirow{3}{*}{3} & F-E  & 0.35 & 0.44 & 0.46 & \textit{\textbf{0.70}} \\ \cline{2-6} 
		& P-S  & 0.37 & 0.39 & 0.50 & \textit{\textbf{0.70}} \\ \cline{2-6} 
		& R-U  & 0.15 & 0.30 & 0.27 & \textit{\textbf{0.42}} \\ \hline
		\multirow{3}{*}{4} & F-E  & 0.51 & 0.43 & 0.44 & \textit{\textbf{0.67}} \\ \cline{2-6} 
		& P-S  & 0.21 & 0.23 & 0.25 & \textit{\textbf{0.40}} \\ \cline{2-6} 
		& R-U  & 0.53 & 0.52 & 0.55 & \textit{\textbf{0.73}} \\ \hline
		\multirow{3}{*}{5} & F-E  & 0.49 & 0.54 & 0.50 & \textit{\textbf{0.86}} \\ \cline{2-6} 
		& P-S  & 0.31 & 0.37 & 0.40 & \textit{\textbf{0.65}} \\ \cline{2-6} 
		& R-U  & 0.44 & 0.49 & 0.54 & \textit{\textbf{0.83}} \\ \hline
		\multirow{3}{*}{6} & F-E  & 0.63 & 0.69 & 0.73 & \textit{\textbf{0.89}} \\ \cline{2-6} 
		& P-S  & 0.25 & 0.29 & 0.34 & \textit{\textbf{0.53}} \\ \cline{2-6} 
		& R-U  & 0.66 & 0.65 & 0.55 & \textit{\textbf{0.74}} \\ \hline
	\end{tabular}
\end{table} 

\subsection{Inter-session Estimations in Single/Multiple DoFs Tasks}
 Fig. 8 illustrates the inter-session performance of CNN and CNN-LSTM in P1-P3 of Subject 5. Performances of both CNN and CNN-LSTM become worse compared to intra-session evaluations in Fig. 6 due to domain shifts among different sessions, but the curves reconstructed by CNN-LSTM still manage to match the ground-truth with promising smoothness and precision. Fig. 9 illustrates comparisons among DL and ML techniques following P1-P4. In terms of $R^{2}$ values, CNN-LSTM outperforms other techniques significantly, particularly in P4. As for wrist motions in flexion and extension, $R^{2}$ values of CNN-LSTM can be as high as 0.93 and 0.74 in new testing sessions of P1 and P4 (F-E), respectively, indicating a reliable proportional myoelectric control in this DoF. In fact, promising accuracies can be achieved by ML models in P1 (SVR and RF reach 0.73 and 0.79, respectively). As is discussed in visual exploration, the high accuracies in F-E benefit from the upper limb anatomy, which on the other hand leads to non-negligible cross-talk for sEMG of other DoFs in P4. 
 
 \begin{figure}[!t]
 	\centering
 	\setlength{\belowcaptionskip}{-0cm}   
 	\includegraphics[width=3.5in]{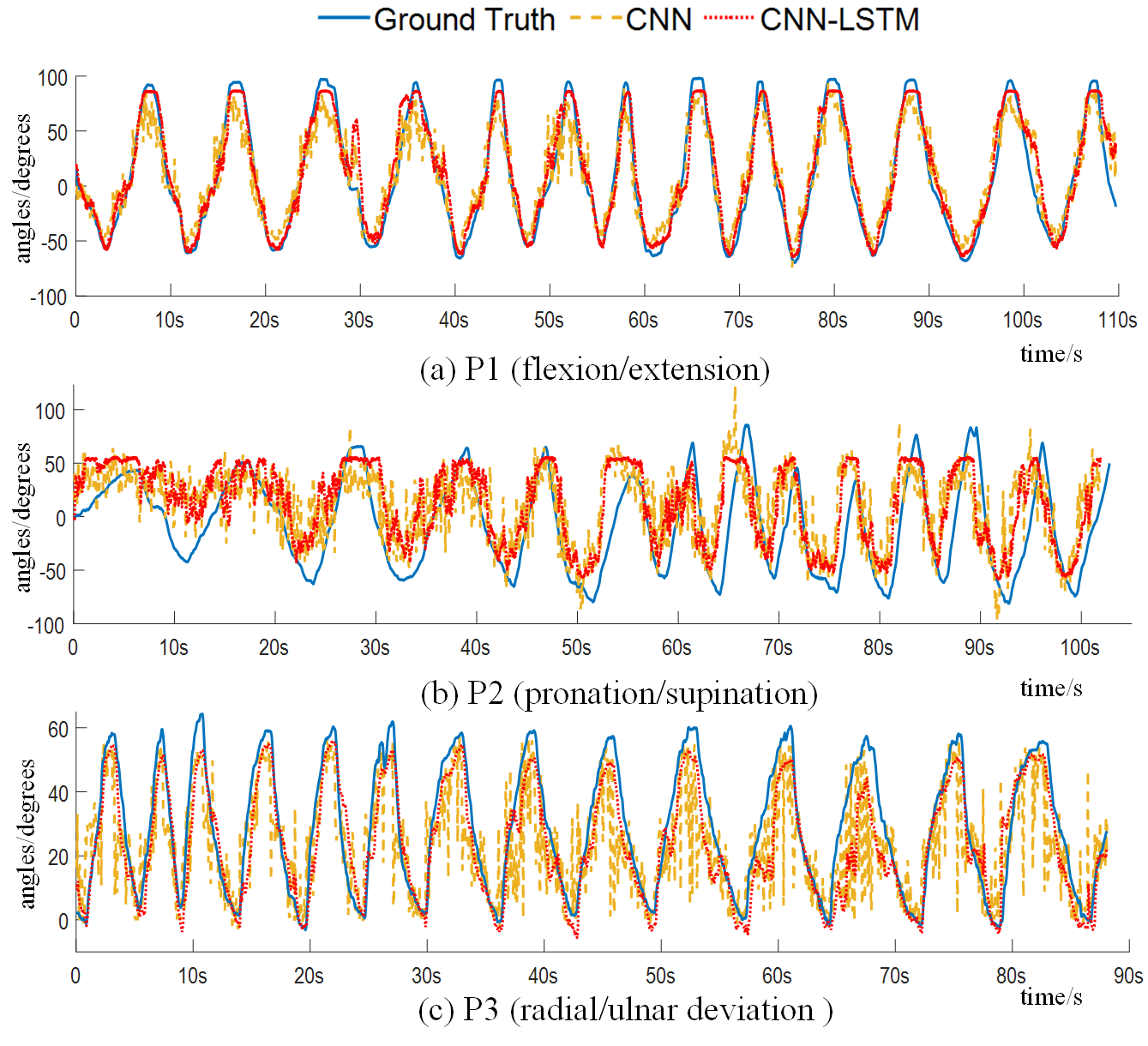}
 	\caption{Inter-session estimations of CNN and CNN-LSTM following P1-P3 of Subject 5. (a) P1(F-E). (b) P2(P-S). (c) P3(R-U).}
 	\label{Inter-session estimations}
 \end{figure}

\begin{figure}[!t]
	\centering
	\vspace{0cm}  
	\setlength{\belowcaptionskip}{-0cm}   
	\includegraphics[width=3.5in]{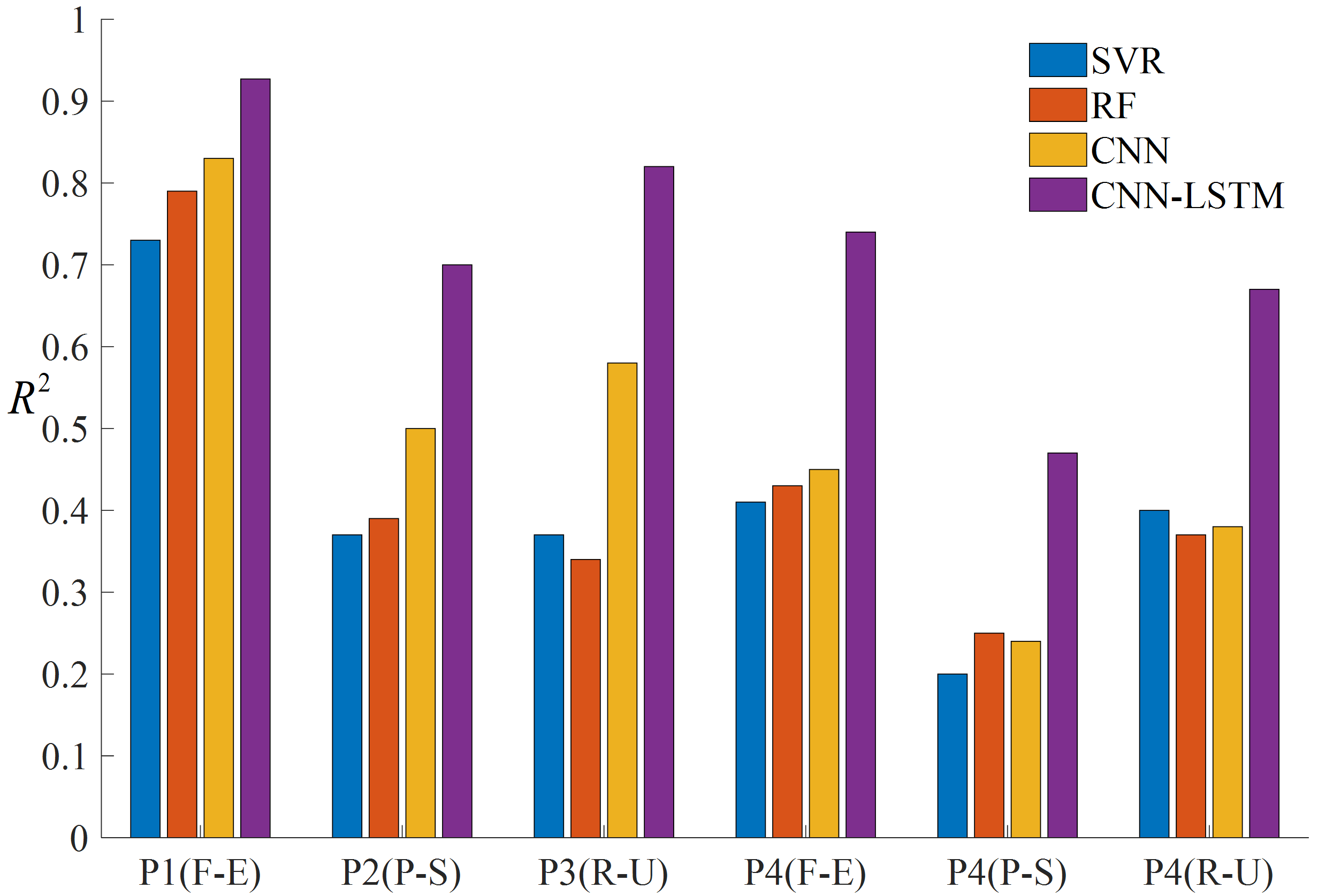}
	\caption{Inter-session evaluations of SVR, RF, CNN and CNN-LSTM in P1-P4.}
	\label{evaluations of SVR, RF, CNN and CNN-LSTM}
\end{figure}

In spite of the dramatic increase in estimation accuracies compared with several representative regression techniques, the performance of CNN-LSTM is not always encouraging (lower than 0.5 for $R^{2}$ values) in very tough tasks such as P4 (P-S) in inter-session evaluations. As verified in previous studies, wrist kinematics estimations can be very challenging with more than two-DoFs involved \cite{amsuess2015context}. It is reasonable since the regression task requires high resolution of data but sEMG indeed suffers a lot from its property of a low signal-to-noise ratio (SNR). Therefore, further improvements are still yielded in either model construction, data acquisition, signal processing and experiment protocols design, etc. Only with accurate, robust and efficient wrist kinematics estimations, can simultaneous and proportional myoelectric control be indeed accomplished in practical applications.
 
\subsection{Comparison of Time-steps in CNN-LSTM}
Hyper-parameters are of vital importance to DL techniques. Compared with optimizer hyper-parameters such as learning rate, batch size and iteration process, the time-step $k$ in the rearranged feature sequence $\left[\mathbf{f}_{1}, \mathbf{f}_{2} \cdots \mathbf{f}_{k}\right]$ is more task-specific. It determines the number of samples in sequence regression, thus a larger $k$ denotes a longer term of time dependencies which may contribute to a higher accuracy but also results in a heavier computational load. In this subsection four different time-steps are primarily evaluated for CNN-LSTM, i.e. 8, 18, 58, 98 for the value of $k$ which correspond to 0.5s, 1s, 3s and 5s in time duration, respectively.

\begin{figure}[!t]
	\centering
	\vspace{-0.2cm}  
	\includegraphics[width=3.5in]{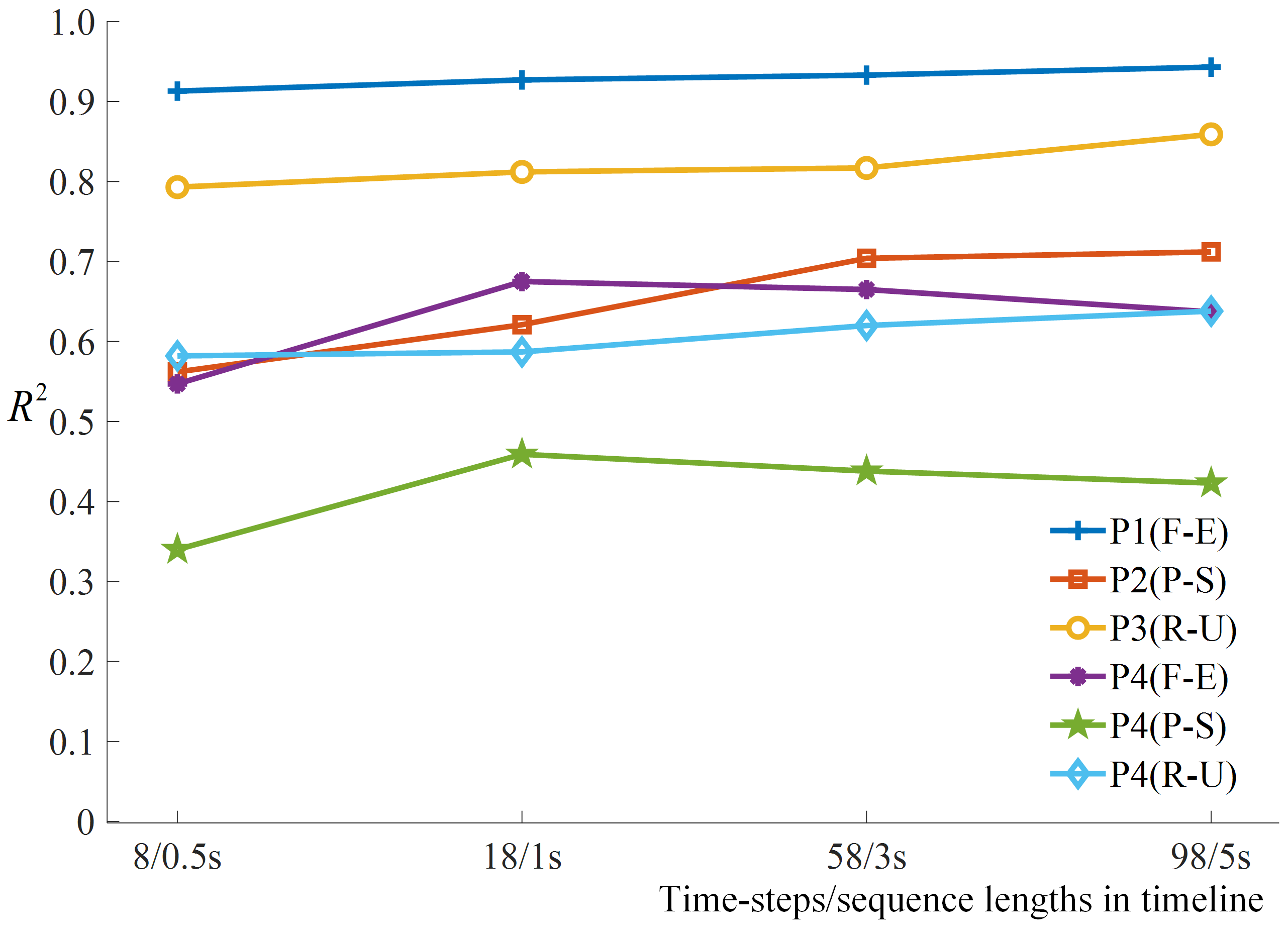}
	\caption{Comparison between time-steps/sequence lengths of CNN-LSTM in inter-session evaluations. }
	\label{different time-steps}
\end{figure}

Estimation results of LSTM with different time-steps in inter-session evaluations are illustrated in Fig. 10. In general, the $R^{2}$ value of CNN-LSTM improves gradually along with the increase of time-steps. It can be inferred that the exploitation of long-term time dependencies contributes to a higher estimation accuracy in most scenarios. However, this benefit varies among test protocols and may lead to over-fitting which can be found in P4 (F-E) and P4 (P-S). Empirically, a sequence in 1s duration can reach a compromise in model effectiveness and efficiency. Besides, a too large sequence is inapplicable for real-time myoelectric control since the intention prediction is expected to be implemented without evident time delays. 

\begin{figure}[!t]
	\centering
	\vspace{-0cm}  
	\includegraphics[width=3.5in]{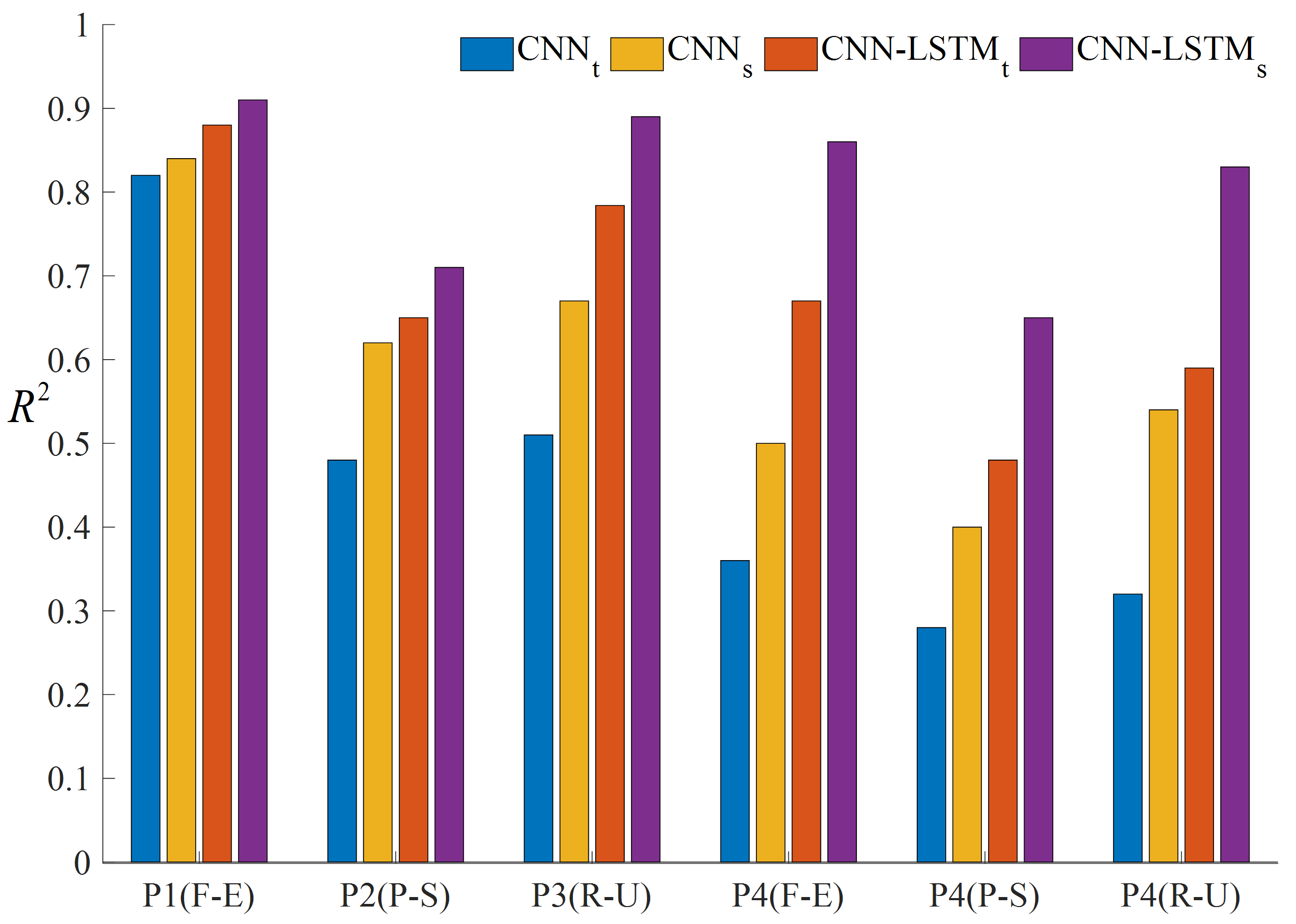}
	\caption{Intra-session evaluations of CNN and CNN-LSTM with two types of sEMG matrices. CNN/CNN-LSTM with temporal or spectral inputs are shorted as CNNt, CNNs, CNN-LSTMt and CNN-LSTMs, respectively.}
	\label{different time-steps}
\end{figure}

\subsection{Comparison of sEMG Matrices}
Besides the architecture and hyper-parameters, sEMG input matrices also have a non-negligible impact on CNN-based feature extraction and can then influence the estimation accuracy of CNN-LSTM. In Section II, we obtain the spectrum-based sEMG matrices by applying FFT on each sliding window. A more intuitive method is to construct matrices in the time domain directly. The comparison of CNN and CNN-LSTM with temporal and spectral sEMG matrices in intra-session evaluations can be found in Fig. 11. For simplicity, CNN/CNN-LSTM with temporal or spectral inputs are shorted as CNNt, CNNs, CNN-LSTMt and CNN-LSTMs, respectively. It can be observed that CNNs outperforms CNNt in all protocols, which contributes to the outperformance of CNN-LSTMs over CNN-LSTMt accordingly. This superiority becomes more significant in multi-DoF tasks. A possible reason is that the sEMG collected by sparse electrodes can be regarded as the superimposition of signals from multiple muscles. During voluntary contractions, the firing rates of motoneuron in these muscles are different \cite{seki1996firing}, thus the spectrum information can be more representative and distinguishable.

\section{Conclusion and Future Work}

In this paper, we presented a hybrid framework to combine CNN-based feature extraction and LSTM-based sequence regression in wrist kinematics estimation, which could extract temporal-spatial correlations in sEMG efficiently. Through visual exploration, we verified that deep features extracted by CNN were more representative than traditional hand-crafted features. By exploiting contextual information in a deep feature sequence, the presented CNN-LSTM outperformed conventional CNN, SVR and RF significantly in both intra-session and inter-session evaluations, particularly when wrist
movements were activated in multi-DoFs.

Because of the non-stationary characteristics of sEMG, our future research will focus on longitudinal/multiday evaluations of ML and DL techniques in wrist kinematics estimation. Since data-driven methods rely on the assumption that training and testing data should stem from same underlying distributions, it is essential to further investigate domain/rule adaptation approaches suitable for the CNN-LSTM model. Besides, to support deep learning in wearable systems, we will also work on the quantization method of our hybrid framework and its implementations using FPGA accelerator. 

\footnotesize

\bibliographystyle{ieeetran}
\bibliography{CNN-LSTMRef}

\end{document}